# A new triclinic modification of the pyrochlore-type $KOs_2O_6$ superconductor

S. Katrych[1]*, Q.F. Gu[2], Z. Bukowski[1], N.D. Zhigadlo[1], G. Krauss[2], J. Karpinski[1]

[1]*Laboratory for Solid State Physics, Schafmattstr. 16, ETH Zürich, 8093 Zürich, Switzerland*

[2]*Laboratory for Crystallography, Wolfgang-Pauli-Str. 10, ETH Zürich, 8093 Zürich, Switzerland*

**Abstract**

A new modification of $KOs_2O_6$, the representative of a new structural type (Pearson symbol a$P$18, $a$=5.5668(1) Å, $b$=6.4519(2) Å, $c$=7.2356(2) Å, $\alpha$=65.377(3)°, $\beta$=70.572(3)°, $\gamma$=75.613(2)° space group $P$-1, no.2 was synthesized employing high pressure technique. Its structure was determined by single-crystal X-ray diffraction. The structure can be described as two $OsO_6$ octahedral chains relating to each other through inversion and forming big voids with K atoms inside. Quantum chemical calculations were performed on the novel compound and structurally related cubic compound. High-pressure X-ray study showed that cubic $KOs_2O_6$ phase was stable up to 32.5(2) GPa at room temperature.

## 1.Introduction

The $\beta$-pyrochlore osmium oxides $AOs_2O_6$ (A = Cs, Rb, K) are highly interesting compounds due to their fascinating properties. They are superconductors at the transition temperatures ($T_c$) of 3.3, 6.3 and 9.6 K respectively [1, 2, 3]. The structure of these compounds consists of $OsO_6$ octahedral framework that forms big voids where alkali metal ions are located [4, 5]. By substitution of alkaline metal in the voids it is possible to tailor physical properties of pyrochlore compounds.

Particular attention was devoted to the $KOs_2O_6$ pyrochlore with the highest $T_c$. In this compound the K ions have huge displacement parameters [4]. Several of interesting physical phenomena, such as second order phase transition in superconducting state [6, 7], unusual temperature dependence of the electrical resistivity [8], uncommon pressure dependence on $T_c$ [9], were associated with the active movement (rattling) of the alkali metal ion in the big voids [10].

---
[*] katrych@phys.ethz.ch , Fax: +41 44 633 1072

During the study of anomalous pressure dependence of magnetization in the cubic $KOs_2O_6$ superconductor it was found that $T_c$ slightly increases with increasing pressure (up to 0.56 GPa), become saturated (at 1.2 GPa) and then falls down [9]. The superconductivity vanishes at about 6 GPa [11]. Structural investigations at high pressure (HP) are crucial for a better understanding of the superconducting properties of $AOs_2O_6$ compounds on pressure. By using HP for crystal growth it was possible to modify the cubic structure of $KOs_2O_6$ and to obtain a HP modification of $KOs_2O_6$.

We report on the synthesis, crystal structure and quantum chemical calculations on this new compound as well as HP powder x-ray diffraction study of the cubic $KOs_2O_6$.

## 2. Experimental

*2.1. Preparation*

Powders of $KO_2$, $OsO_2$ and $AgO$ were well mixed and pressed in the Au cylindrical capsule with a diameter 6.8 mm. AgO was added as a source of oxygen. The capsule was placed in to BN crucible and finally they were mounted in a pyrophillite cube. A graphite tube was used as a heating element. A set of steel parts transmitted force through six tungsten carbide pistons to the sample in a quasi isostatic way. A pressure up to 30 kbar was applied at room temperature. While the pressure was kept constant, the samples were heated during 1.5 h up to the 900 °C and kept at this temperature for 1 h. After that the temperature was slowly ramped to 500 °C within 2 h, held at 500 °C for 1 h and finally the sample was cooled to the room temperature during 1 h.

The cubic $KOs_2O_6$ was synthesized from stoichiometric amount of Os, $KO_2$ mixture as the method described by [12].

*2.2 Single crystal x-ray diffraction and magnetization measurements on triclinic $KOs_2O_6$*

After crushing the lump a large number of well shaped crystals with the sizes of a few hundreds μm were found (Fig. 1). Several selected single crystals were investigated on a single-crystal X-ray diffractometer equipped with CCD detector (Xcalibur PX, Oxford Diffraction) with 60 mm sample-detector distance. For further details on data collection see Table 1. Data reduction and gaussian absorption correction were performed using the software package CrysAlis [13]. The crystal structure was solved by Patterson method [14] and the full data set refined on $F^2$, employing the programs SHELXS-97 and SHELXL-97 [15] and [16]. Chemical composition was determined using energy-dispersive X-ray spectroscopy (EDX).

Magnetization measurements have been performed on the collection of nonoriented single crystals using a SQUID magnetometer.

*2.3 High Pressure studies on cubic $KOs_2O_6$*

In order to understand the crystal structure behavior under pressure and structure relationship between triclinic phase and cubic phase, we studied the cubic $KOs_2O_6$ under HP. Cubic $KOs_2O_6$ single crystals were selected and ground into powder. The lattice parameters were determined as $a$ = 10.095(1) Å on a STOE-diffractometer (Co-Kα1 radiation) with silicon (NBS640a) as internal standard. HP synchrotron x-ray diffraction experiments were performed at the Materials Science beam line, Swiss Light Source (SLS) at Paul Scherrer Institute (PSI, Villigen, Switzerland). The 2D diffraction data was collected on a Marresearch mar345 image-plate detector using a wavelength of 0.5640 Å. The 2D diffraction images were integrated by the use of the program fit2d [17] and the 1D powder pattern were refined by applying the Le-Bail fit using the program GSAS [18, 19]. High pressures up to 32.5(2) GPa were generated by use of an ETH-type diamond anvil cell at room temperature. A mixture of methanol and ethanol (4:1) was served as pressure-transmitting medium. To minimize deviatoric stress, the amount of pressure medium was covered ca. 30% of the total volume (materials 70%) of the sample chamber within tungsten gasket. A ruby crystal was placed into the sample chamber as a pressure marker and the pressure was determined using the ruby fluorescence technique. Equation of state was calculated by the use of the program EosFit5.2 [20].

*2.4 Quantum chemical calculations*

Quantum chemical calculations were performed by use of the Vienna *ab-initio* simulation package (VASP) [21, 22] based on the projected-augmented wave function (PAW) method [23]. The structures were geometry optimized in VASP, which employs ultrasoft pseudopotentials [24] and the generalized gradient functional (GGA) to model the contribution of the exchange correlation to the total energy and the valence electron density distribution. The kinetic energy cutoff for the runs was chosen to be sufficient to ensure convergence of the energy. Optimization of the structural parameters was performed until the forces on the atoms were less than 0.01 eV/Å and all stress components were below 0.001 eV/Å$^3$. The electron localization function (ELF) was evaluated according to [25] with an ELF module implemented in VASP. Bader analysis of charge density of atoms was done according to Bader volume [26] by using Bader code [27].

## 3. Results and discussion

*3.1 Structure analysis of triclinic phase*

After statistical analysis of normalized structure factors the centrosymmetric space group was chosen from two possible variants *P*-1 and *P*1. The atomic positions of Os and K atoms were found by Patterson method. The structure model obtained was refined without any constrains. The results of the structure determination are presented in Tables 2 - 5.

The Os sublattice is constituted from two zigzags like chains along the *b* direction related to each other through inversion (Fig. 2a). The Os1-Os2-Os1 angle is 130.792(4) deg. The K atoms are located between Os layers (Fig. 2a). The O atoms build octahedra with the Os atom in the centre. The $OsO_6$ octahedra are connected via common corners and present two chains along the Os zigzags. It is very interesting that, as in the cubic $KOs_2O_6$ - pyrochlore structure [4], the structure of triclinic $KOs_2O_6$ involves a network with the same $OsO_6$ octahedra as the building blocks and K atoms are located in the big voids (Fig. 3 and Fig. 2bc). There are two types of $OsO_6$ octahedra $Os(1)O_6$ and $Os(2)O_6$ (Fig. 4). They are much more distorted than octahedra in the cubic $KOs_2O_6$ (Tab. 4 and [4])

The unit cell of HP triclinic $KOs_2O_6$ is about 17 % more dense than unit cell of the cubic prototype. In comparison to the cubic modification the differences between atomic displacement parameters $U_{iso}$ are not so large. For example, $U_{iso}$ for the K atom in cubic pyrochlore 17-14 time larger than $U_{iso}$ for Os [4]. The $U_{iso}(K)/U_{iso}(Os)$ ratio for the triclinic $KOs_2O_6$ is only about 2 (Table 2). That could be explained with the decreasing of the free space between $OsO_6$ octahedra and with the considerably shorter K-K distance (K-$K_{triclinic}$ = 3.247(2) Å, K-$K_{cubic}$ ~ 4.4 Å, Fig. 2, 3, Tab. 5). The main structural block, distorted $OsO_6$ octahedra seems to be very stable and remains in the HP modification (the edges O-O distances vary from 2.392(3) to 2.980(5) Å). Further details of the crystal structure investigation can be obtained from the Fachinformationszentrum Karlsruhe, 76344

Eggenstein-Leopoldshafen, Germany, (fax: (49) 7247-808-666; e-mail: crysdata@fiz.karlsruhe.de) on quoting the depository number CSD419880.

According to an EDX analysis the ratio K/Os = 0.51 agrees with crystallographic formula where K/Os = 0.5. Except K, Os, O elements, no other chemical elements were detected. Pearson symbol (PS) of the new compound is $aP18$ and space group $P$-1. There are a lot of compounds with such PS and space group [33], but there are no compounds with lattice dimensions and atomic arrangements like in HP $KOs_2O_6$. We conclude that HP $KOs_2O_6$ is the first representative of a new structural type.

*3.2 HP study on cubic phase*

In order to study the possible phase transition from cubic phase to triclinic phase of $KOs_2O_6$, HP X-ray diffraction was performed on cubic phase. Fig. 5 shows some representative powder diffraction patterns obtained from synchrotron measurements. Except the broadening, no drastic structural changes of the diffraction pattern were observed up to the highest reached pressure at 32.5(2) GPa. All the diffraction peaks can be indexed by cubic $KOs_2O_6$, and Os being present as an impurity. After pressure release, the observed patterns are comparable to those collected before the high-pressure study. A 3rd order Birch-Murnaghan equation of state was fitted to the experimental data (Fig. 6). The values of the bulk modulus based on third-order Birch-Murnaghan equations-of-state are $K_0$=154(4) GPa with $K′$=3.3(2). The cubic structure was stable within the framework of this experiment. This is surprising, as during the HP synthesis with 3 GPa and 900 °C only the triclinic phase was produced. The calculated enthalpies per formula unit for both cubic and triclinic $KOs_2O_6$ are very close at ambient pressure (Fig. 7). With increasing pressure, the triclinic structure shows a lower enthalpy compared with the cubic structure, which means triclinic phase is more energy favorable to be a stable phase. The reason that cubic phase can not transform into triclinic phase is possibly due to high energy barrier between two structures, that need to be overcame by high temperature.

*3.3 Bonding*

The calculated electronic density of states (Fig. 8) shows non-zero values at the Fermi level ($E_F$) for both compounds. For both structures, a local maximum is present close to $E_F$. The result of total DOS of the cubic structure is similar with literature data [28-30]. The overall profile of total DOS is similar for both cubic and triclinic band structure. In cubic structure, there are many sharp peaks coming from high symmetry of the beta-pyrochlore structure. The projected DOS indicate that there are nearly no K states near Fermi level in both structures, the electronic structure near the Fermi energy is dominated by strongly hybridized Os and O states. The both structures are dominated by corner shared Os-O octrahedra.

To analyse the chemical bonding of the materials, the electron localization function (ELF) was calculated. ELF is based on the Hartree-Fock pair probability of parallel spin electrons and can be calculated in density functional theory from the excess kinetic energy density due to Pauli-repulsion [25, 31]. This function produces easily understandable, pictorially informative patterns of chemical bonding and is widely used to describe and visualize chemical bonding in molecules and solids [32]. The distribution of ELF in both structures shows the maxima located on the outer side of the Os-O octahedra close to O atoms (Fig. 9). There are small attractors in the Os-O octahedra, between Os and O atoms. This allows interpreting them as strongly hybridized Os and O valence electrons in both structures, as covalent bonding. The Os-O octahedra form interconnected chains through corner O atoms in both structures. The valence shell of the potassium atoms in the crystal structure of both cubic and triclinic is nearly spherical. No additional attractors were found in vicinity of K atoms.

This suggests there is no direct ionic bonding between K atoms to rest of the structures, the K atoms are only loosely sit in the voids of structures. The Bader analysis shows valence state of K(0.92+), Os(2.43+), O(0.96-) in the cubic structure, and K(0.89+), Os(2.22+; 2.38+), O(0.92-; 0.89-; 0.91-; 0.97-) in the triclinic structure, which are quite similar in both structures (Table 6).

*3.3 Magnetization measurements*

Magnetization measurements for the triclinic $KOs_2O_6$ showed temperature independent diamagnetic behaviour above 100 K. At lower temperatures an upturn is observed and sample becomes paramagnetic. Superconductivity has not been detected down to 3 K.

**4. Summary**

The novel compound, triclinic $KOs_2O_6$, reveals close local structural resemblance to the cubic $KOs_2O_6$. Both structures are constructed from the same structural blocks $OsO_6$ octahedra building big voids with K atom inside. K has no direct chemical bonding to the rest of the structure. The quantum chemical calculations show that at HP the triclinic modification is more stable than the cubic one. A high pressure (32.5(2) GPa) is not sufficient factor for transforming the cubic in to the more dense triclinic prototype. High temperature (at least 900 ºC) and a pressure (above 3 GPa) are needed for such possible transformation.

**Acknowledgments**

This work has been supported by the Swiss National Science Foundation under contract numbers 200021-115871 (by Q.F.Gu and G.Krauss) and 116759 (by S. Katrych,

Z. Bukowski, N.D. Zhigadlo, J. Karpinski) as well as by the NCCR program MaNEP. We are grateful to P. Wägli (Electron Microscopy ETH Zurich) for performing of EDS measurements. Part of this work was performed at Swiss Light Source, Paul Scherrer Institute, Villigen, Switzerland. Experimental assistances from the staff of the MS-Beamline are gratefully acknowledged.

*Captions*

Fig. 1. The crystals of triclinic $KOs_2O_6$.

Fig. 2. Schematic representation of the 3x3x3 unit cells projection of triclinic $KOs_2O_6$. From left to the right:
a) Os and K sublattices, the projection along *a* direction before and after applying inversion.
b) structure projection along the *a* direction before and after applying inversion. Black polygon shows big channel where K atoms are located.
c) structure projection along the *b* direction before and after applying inversion.

Fig. 3. Unit cell projection of cubic pyrochlore $KOs_2O_6$. The black polygon shows the big channel where K atoms are located.

Fig. 4. Two types of Os octahedra in triclinic $KOs_2O_6$.

Fig. 5. Selected X-ray powder diffraction patterns of cubic $KOs_2O_6$ at different pressure. Some small amount of Os metal impurity is marked with arrows.

Fig. 6. Unit cell volume of cubic $KOs_2O_6$ as a function of pressure. The solid line corresponds to a 3rd-order Birch-Murnaghan equation of state. The errors are smaller than the size of the symbols.

Fig. 7. Comparison of calculated enthalpy per formula unit as a function of pressure for both cubic and triclinic $KOs_2O_6$.

Fig. 8. Electronic density of states for cubic and triclinic $KOs_2O_6$. The upper figures are total DOS, contributions of Os, O, and K valence electrons are shown as in lower figures as projected density of states. (colour online)

Fig. 9. Electron localization function (ELF $\eta$) for cubic (upper illustration) and triclinic (lower illustration) $KOs_2O_6$: isosurface of $\eta = 0.84$ in both figures, 2D cross section cut through K, Os and O atoms. (colour online)

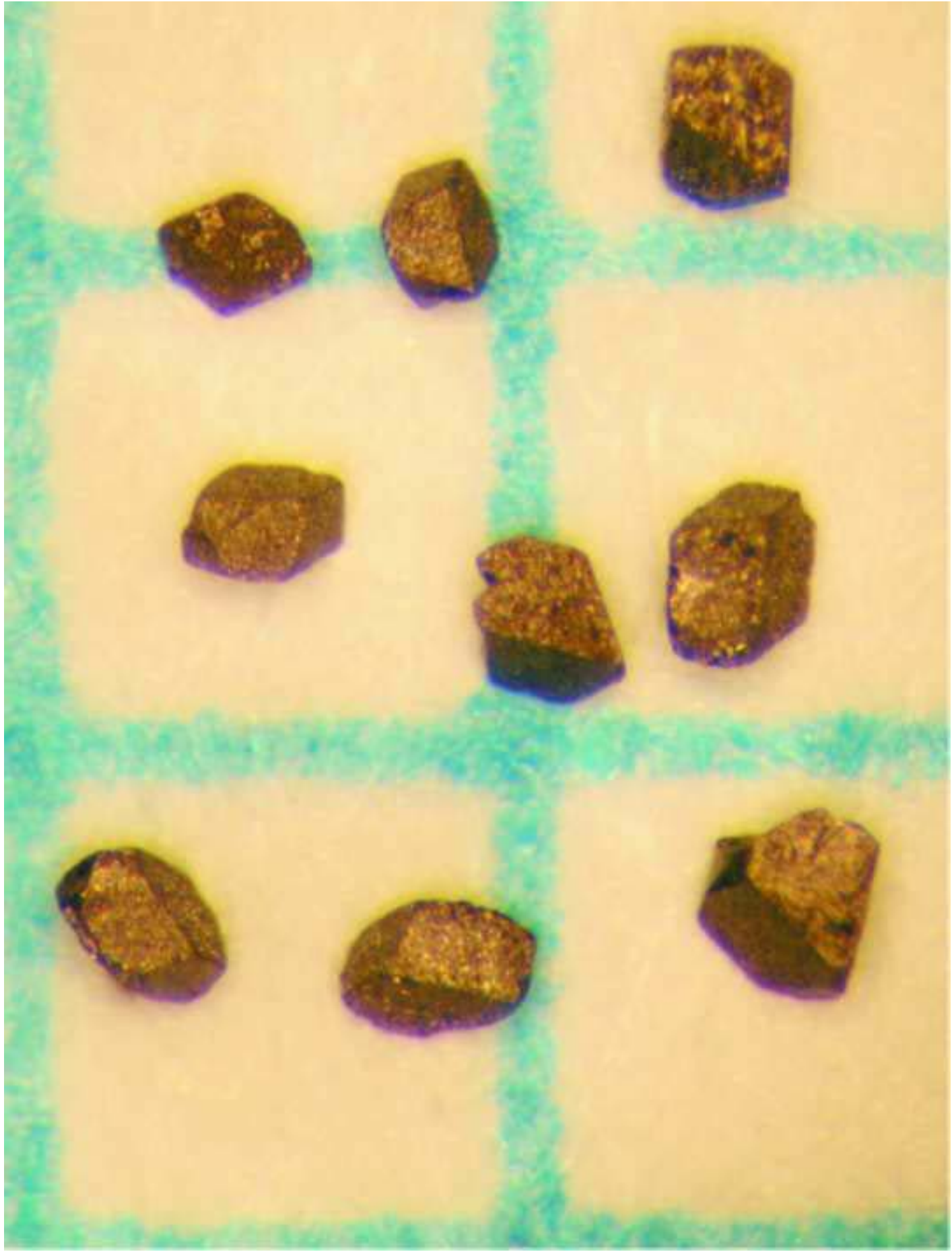
**Figure**
Click here to download high resolution image

**Figure**
Click here to download high resolution image

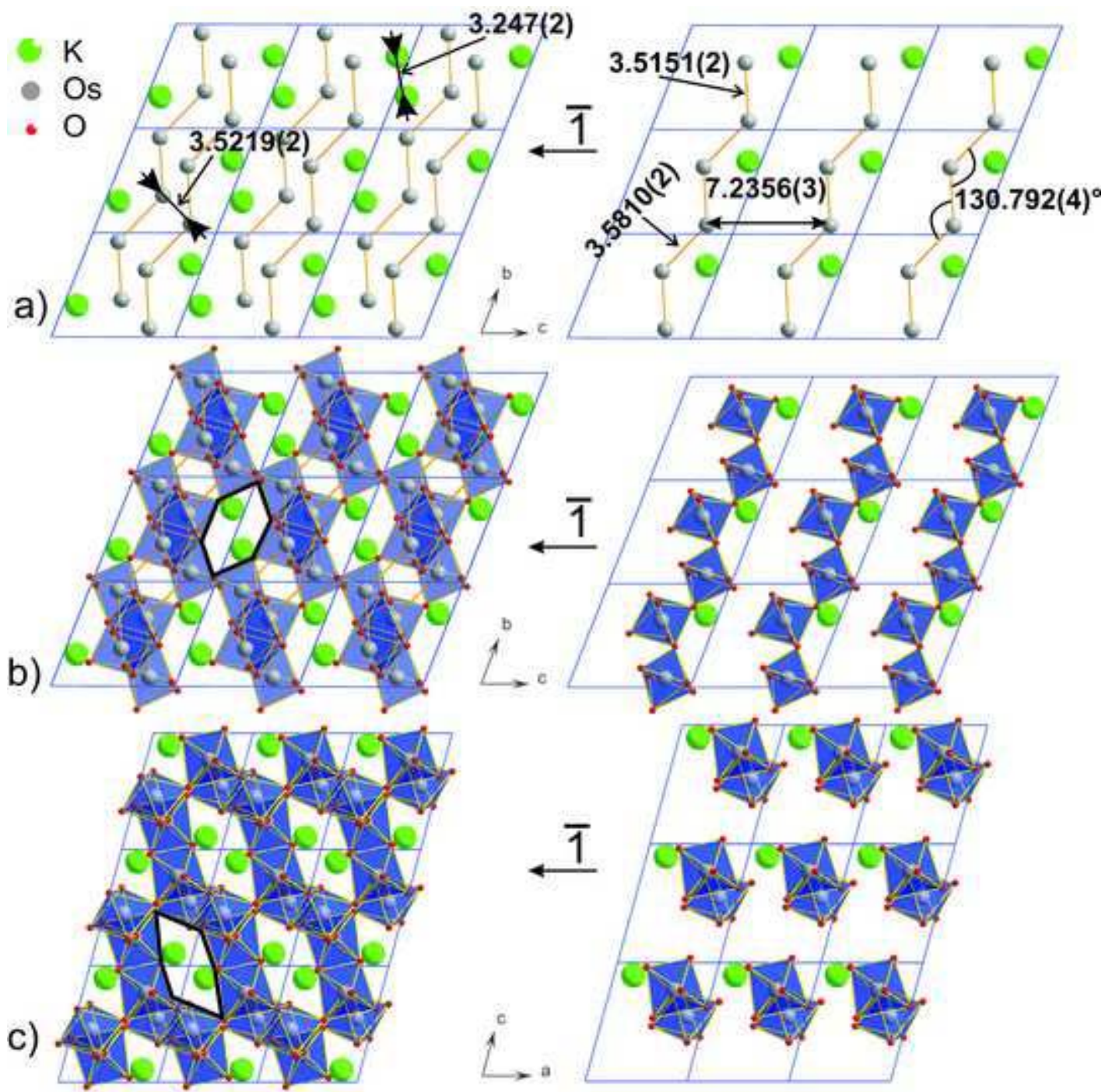

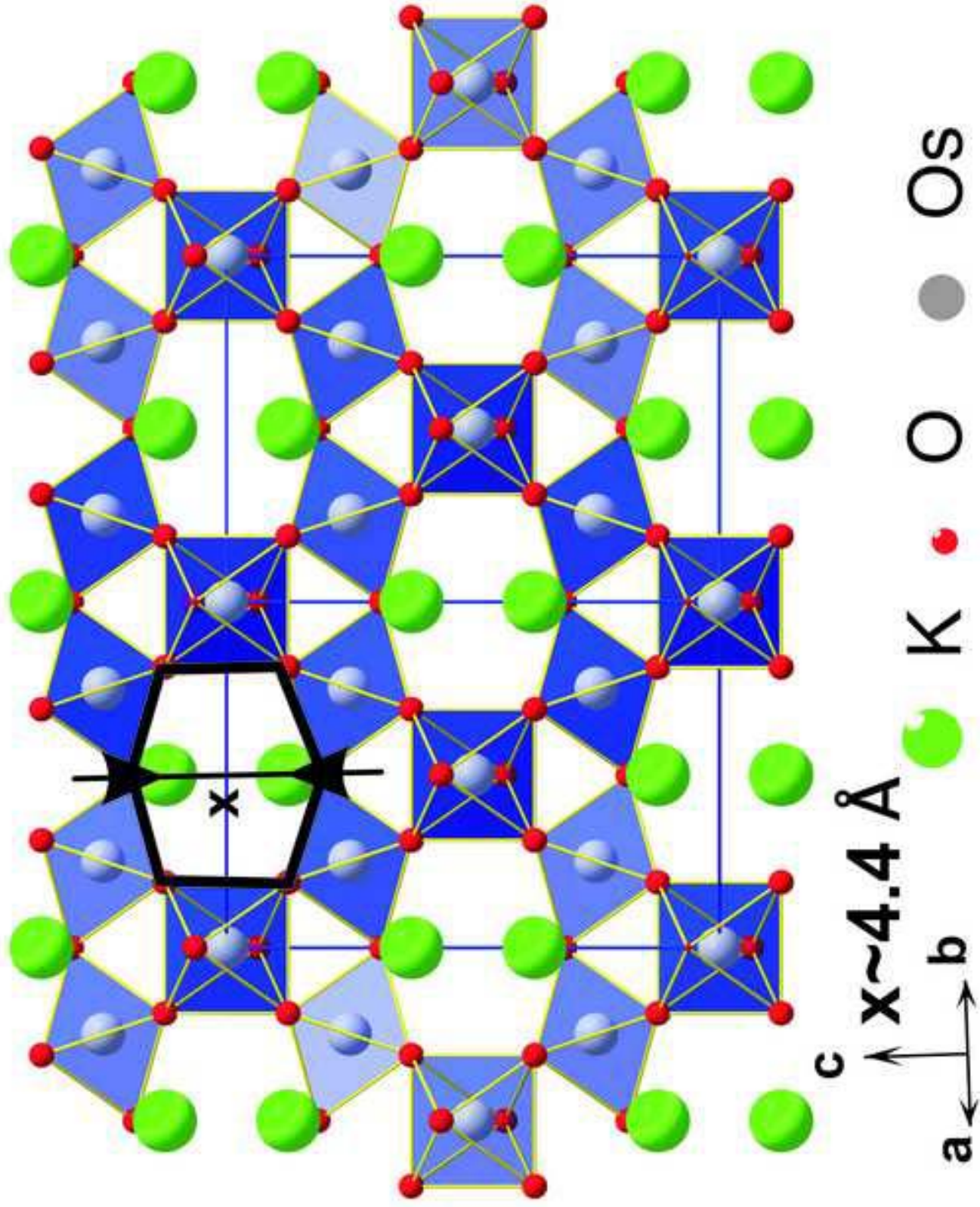

**Figure**
[Click here to download high resolution image](#)

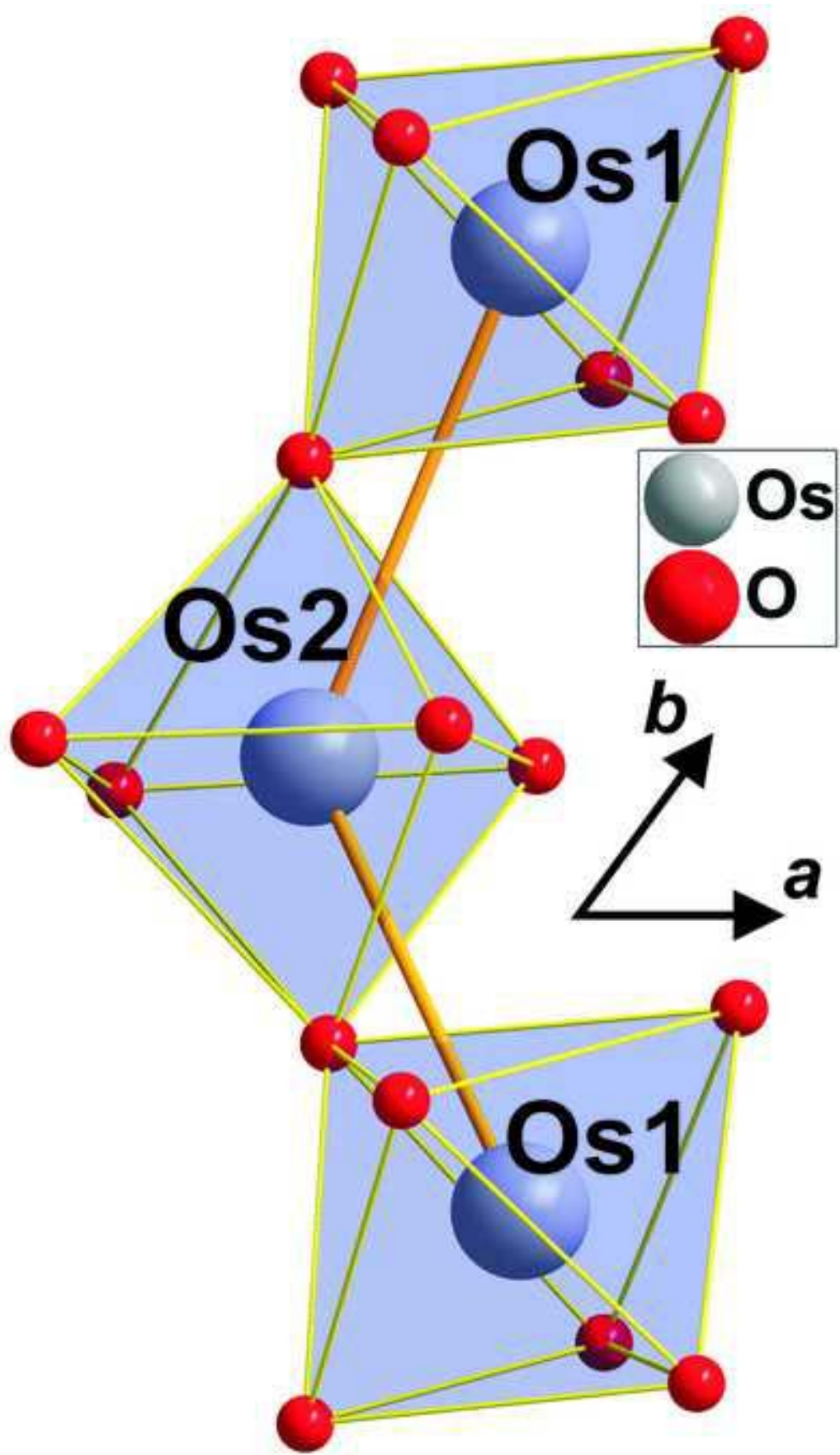

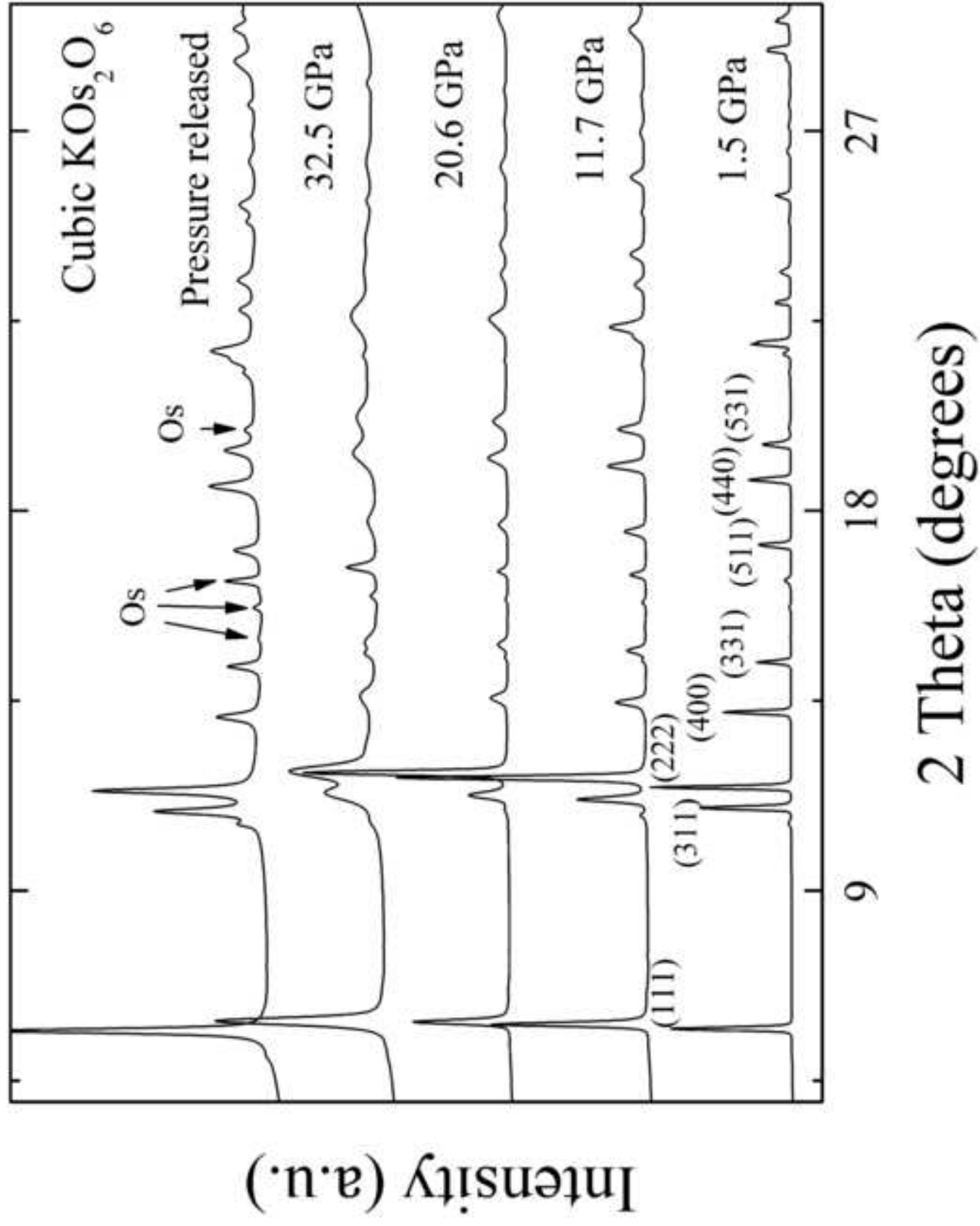

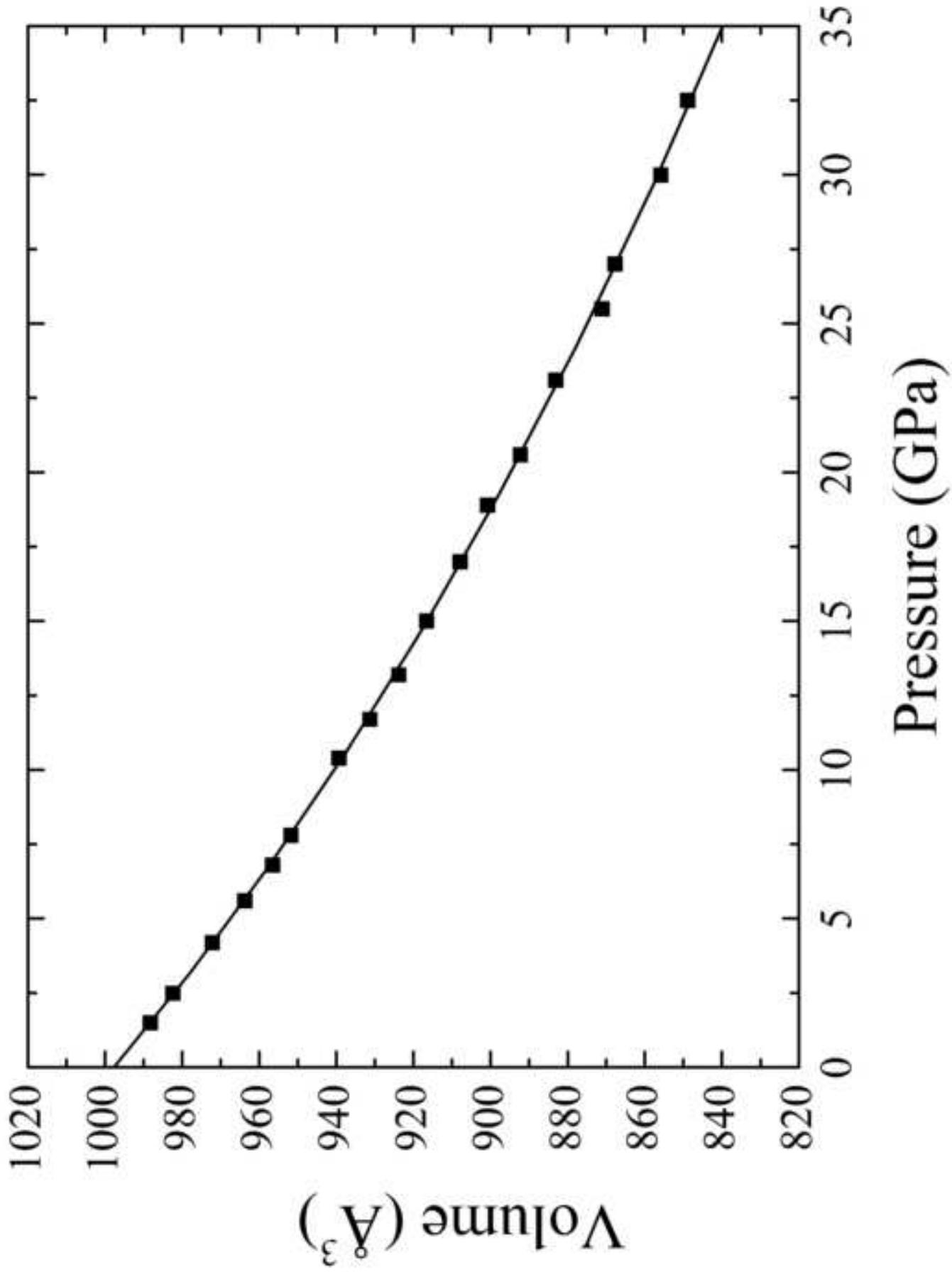

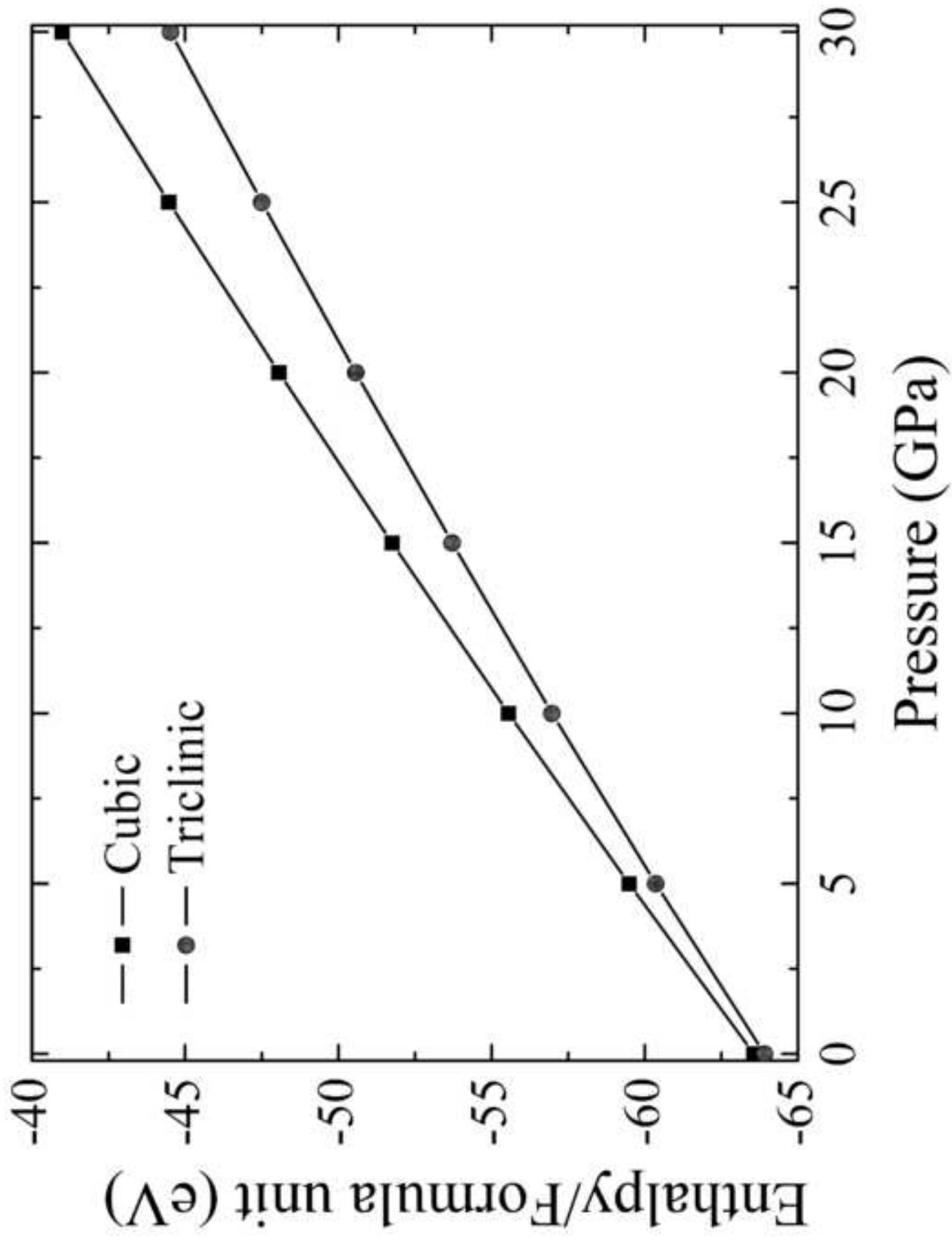

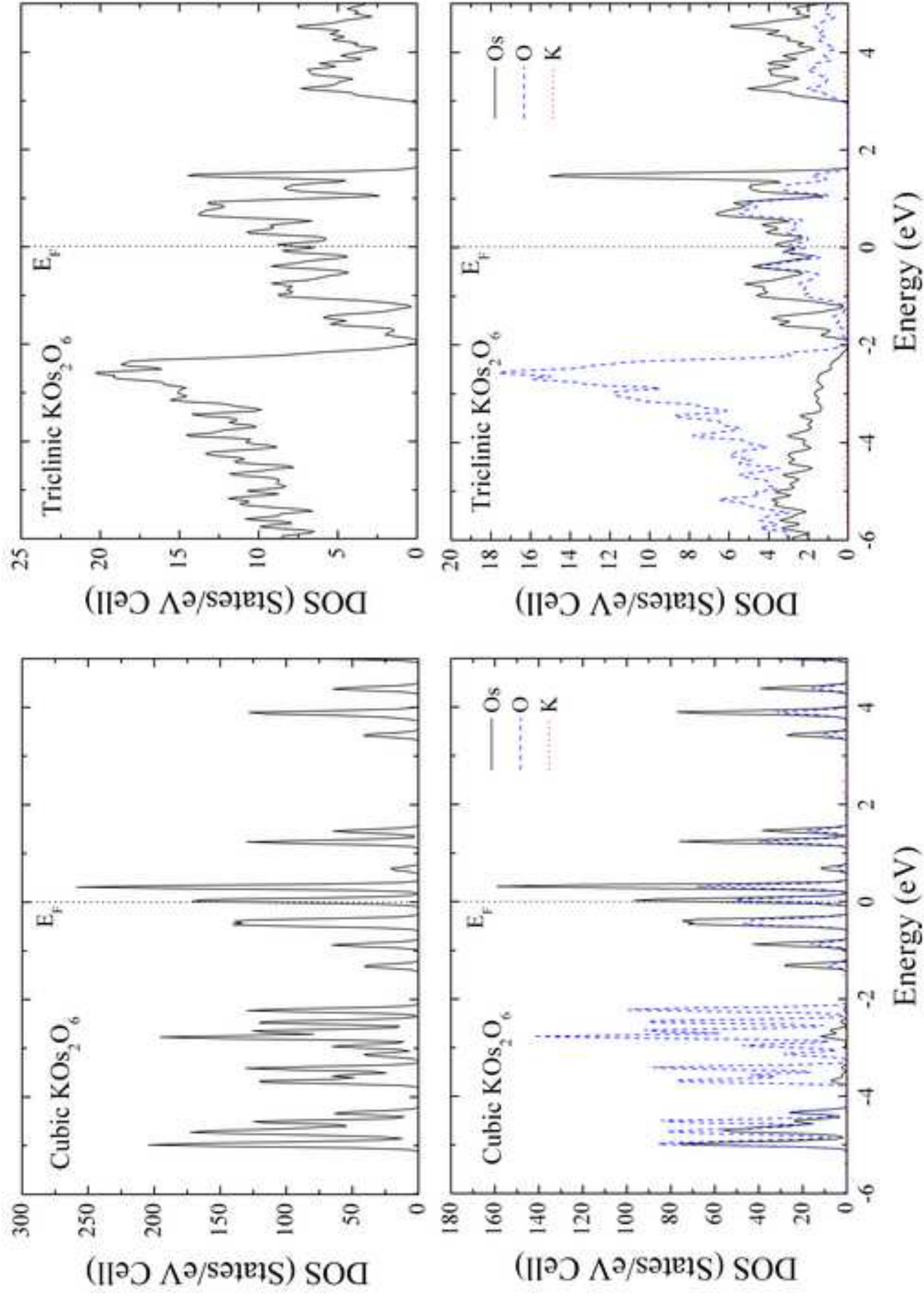

**Figure**
[Click here to download high resolution image](#)

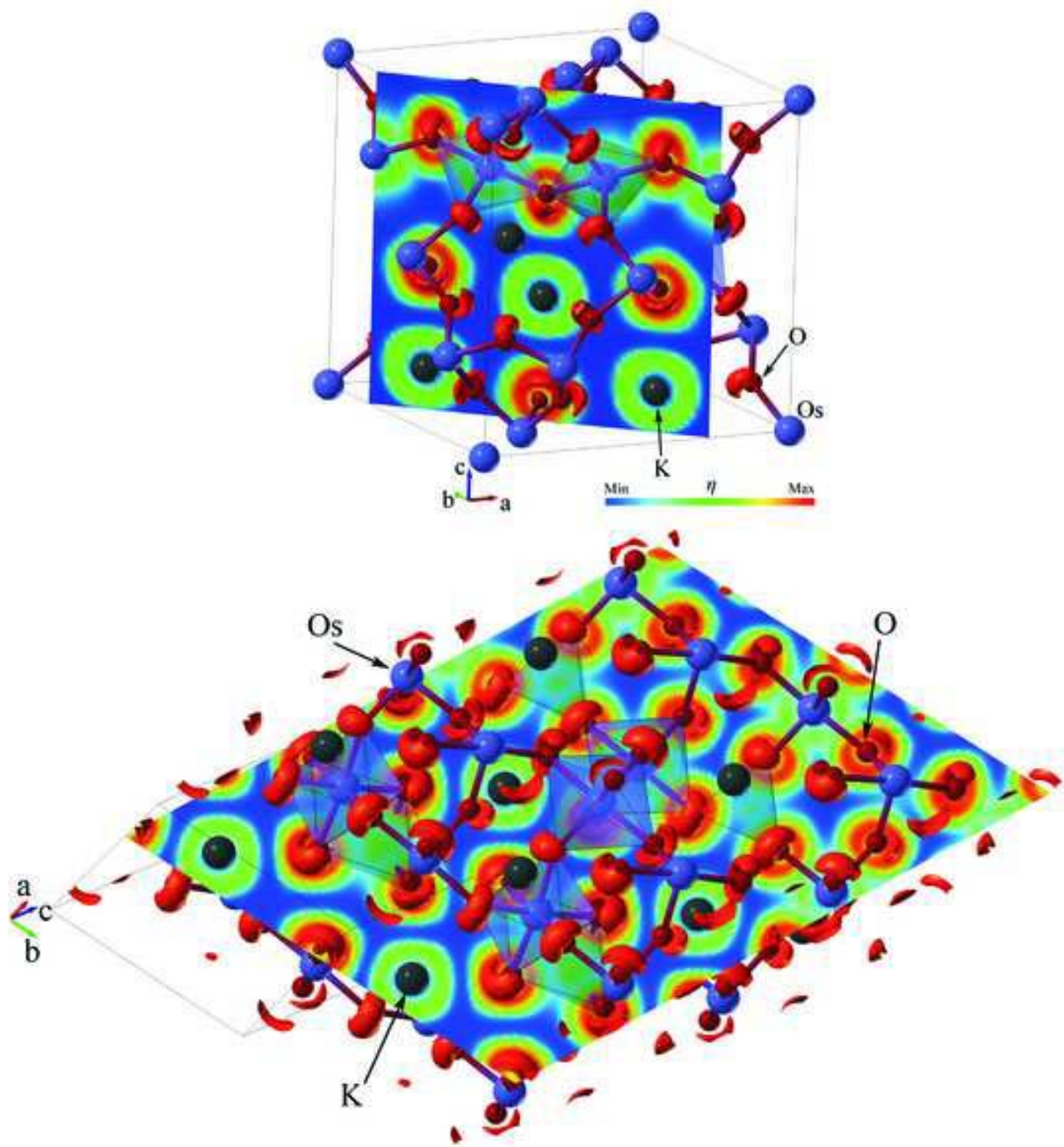

**Table**

Table 1. Crystal data and structure refinement for the triclinic $KOs_2O_6$.

| Empirical formula | $KOs_2O_6$ |
|---|---|
| Formula weight, g/mol | 515.5 |
| Temperature, K | 295(2) |
| Wavelength, Å | 0.71073/MoK$\alpha$ |
| Crystal system, space group, Z | triclinic, $P$-1, 2 |
| Unit cell dimensions, Å, deg | $a$= 5.5668(1), $\alpha$= 65.377(3) |
| | $b$= 6.4519(2), $\beta$= 70.572(3) |
| | $c$= 7.2356(2), $\gamma$= 75.613(2) |
| Volume, Å$^3$ | 220.957(10) |
| Calculated density, g/cm$^3$ | 7.748 |
| Absorption correction type | gaussian |
| Absorption coefficient, mm$^{-1}$ | 58.333 |
| $F(000)$ | 438 |
| Crystal size, mm$^3$ | 0.07 x 0.05 x 0.05 |
| Theta range for data collectio | 3.91 to 42.64 deg. |
| Index ranges | -10<=h<=9, -12<=k<=12, -13<=l<=13 |
| Reflections colected/unique | 8881/3193  $R_{int.}$= 0.0200 |
| Max. and min. transmission | 0.1585 and 0.0895 |
| Completeness to 2theta | 99.6 % |
| Refinement method | Full-matrix least-squares on $F^2$ |
| Data/restraints/parameters | 3193/0/82 |
| Goodness-of-fit on $F^2$ | 1.019 |
| Final R indices [I>2sigma(I)] | $R_1$ = 0.0195, w$R_2$ = 0.0470 |
| $R$ indices (all data) | $R_1$ = 0.0248, w$R_2$ = 0.0480 |
| $\Delta\rho_{max}$ and $\Delta\rho_{min}$,(e/A$^3$) | 3.695 and -4.187 |

Table 2. Atomic coordinates and equivalent isotropic displacement parameters [Å$^2$ x 10$^3$] for the triclinic $KOs_2O_6$.

| Atom | Site | x | y | z | $U_{iso}$ |
|---|---|---|---|---|---|
| Os(1) | 2i | 0.1878(1) | 0.3578(1) | 0.4493(1) | 6(1) |
| Os(2) | 2i | 0.3644(1) | 0.9140(1) | 0.2359(1) | 6(1) |
| K(3) | 2i | 0.1901(2) | 0.6968(1) | 0.8942(1) | 15(1) |
| O(4) | 2i | 0.1040(5) | 0.4796(4) | 0.6727(4) | 9(1) |
| O(5) | 2i | 0.4950(5) | 0.1867(4) | 0.5353(4) | 10(1) |
| O(6) | 2i | 0.2971(4) | 0.0427(4) | 0.9602(3) | 8(1) |
| O(7) | 2i | 0.3816(5) | 0.6079(4) | 0.2546(4) | 9(1) |
| O(8) | 2i | 0.3025(5) | 0.2228(4) | 0.2311(4) | 9(1) |
| O(9) | 2i | 0.0101(5) | 0.8925(4) | 0.3730(4) | 10(1) |

$U_{iso}$ is defined as one third of the trace of the orthogonalized $U_{ij}$ tensor.

Table 3. Anisotropic displacement parameters [Å x $10^3$] for the triclinic $KOs_2O_6$.

| Atom | $U_{11}$ | $U_{22}$ | $U_{33}$ | $U_{23}$ | $U_{13}$ | $U_{12}$ |
|---|---|---|---|---|---|---|
| Os(1) | 6(1) | 5(1) | 6(1) | -2(1) | -2(1) | 0(1) |
| Os(2) | 6(1) | 6(1) | 6(1) | -2(1) | -2(1) | -1(1) |
| K(3) | 16(1) | 16(1) | 16(1) | -6(1) | -4(1) | -4(1) |
| O(4) | 10(1) | 11(1) | 10(1) | -5(1) | -5(1) | 2(1) |
| O(5) | 8(1) | 10(1) | 10(1) | -3(1) | -4(1) | 0(1) |
| O(6) | 8(1) | 10(1) | 6(1) | -3(1) | -2(1) | -2(1) |
| O(7) | 9(1) | 7(1) | 9(1) | -3(1) | 0(1) | -2(1) |
| O(8) | 10(1) | 8(1) | 9(1) | -5(1) | -2(1) | 1(1) |
| O(9) | 8(1) | 10(1) | 9(1) | -2(1) | -1(1) | -2(1) |

The anisotropic displacement factor exponent takes the form: $-2\pi^2 [ (h^2 a^2 U_{11} + ... + 2hka^*b^*U_{12}]$.

Table 4. Slected bond lengths and angles for $OsO_6$ octhedra in triclinic $KOs_2O_6$

| Os(1)O$_6$ polyhedron | | | | Os(2)O$_6$ polyhedron | | | |
|---|---|---|---|---|---|---|---|
| distances, Å | | angles, deg | | distance, Å | | angles, deg | |
| Os1-O9 | 1.920(2) | | | Os2-O5 | 1.868(2) | | |
| Os1-O7 | 1.927(2) | | | Os2-O9 | 1.901(2) | | |
| Os1-O5 | 1.931(2) | O8-Os1-O5 | 86.1(1) | Os2-O7 | 1.903(2) | O6-Os2-O6 | 75.9(1) |
| Os1-O4 | 1.955(2) | O4-Os1-O8 | 86.2(1) | Os2-O8 | 1.923(2) | O7-Os2-O9 | 85.2(1) |
| Os1-O4 | 1.963(2) | O7-Os1-O8 | 88.3(1) | Os2-O6 | 1.942(2) | O8-Os2-O6 | 87.5(1) |
| Os1-O8 | 1.964(2) | O9-Os1-O4 | 88.3(1) | Os2-O6 | 1.948(2) | O9-Os2-O8 | 87.8(1) |
| O5-O8 | 2.658(5) | O8-Os1-O9 | 88.4(1) | O6-O6 | 2.392(3) | O5-Os2-O8 | 89.9(1) |
| O4-O8 | 2.683(3) | O5-Os1-O4 | 88.7(1) | O9-O7 | 2.574(4) | O6-Os2-O5 | 91.0(1) |
| O7-O5 | 2.704(3) | O5-Os1-O7 | 89.0(1) | O8-O9 | 2.652(4) | O9-Os2-O6 | 91.6(1) |
| O9-O4 | 2.706(3) | O4-Os1-O7 | 89.3(1) | O8-O6 | 2.674(5) | O7-Os2-O5 | 92.1(1) |
| O9-O8 | 2.709(3) | O7-Os1-O4 | 91.0(1) | O8-O5 | 2.679(3) | O6-Os2-O7 | 92.1(1) |
| O7-O8 | 2.710(4) | O4-Os1-O9 | 92.5(1) | O5-O7 | 2.714(5) | O8-Os2-O6 | 92.7(1) |
| O4-O5 | 2.717(4) | O9-Os1-O5 | 93.0(1) | O6-O5 | 2.721(3) | O6-Os2-O7 | 94.0(1) |
| O7-O4 | 2.733(4) | O4-Os1-O4 | 99.02(9) | O6-O9 | 2.756(3) | O5-Os2-O9 | 101.6(1) |
| O7-O4 | 2.768(3) | | | O6-O7 | 2.768(3) | | |
| O5-O9 | 2.793(4) | | | O8-O6 | 2.801(3) | | |
| O4-O9 | 2.799(5) | | | O6-O7 | 2.816(4) | | |
| O4-O4 | 2.980(5) | | | O9-O5 | 2.920(4) | | |

Table 5. Slected bond lengths for triclinic $KOs_2O_6$

| | distances, Å | | |
|---|---|---|---|
| Os1-Os1 | 2.5440(2) | K3-O7 | 2.927(3) |
| Os2-Os2 | 3.0679(2) | K3-O6 | 2.933(2) |
| Os1-Os2 | 3.5151(2) | K3-O8 | 3.027(3) |
| Os1-Os2 | 3.5219(2) | K3-O8 | 3.088(2) |
| K3-O6 | 2.702(3) | K3-K3 | 3.247(2) |
| K3-O4 | 2.734(4) | Os1-K3 | 3.5628(8) |
| K3-O9 | 2.775(2) | Os2-K3 | 3.5822(8) |
| K3-O8 | 2.776(3) | Os2-K3 | 3.594(1) |
| K3-O4 | 2.887(2) | Os1-K3 | 3.6844(8) |
| K3-O5 | 2.895(3) | Os2-K3 | 3.766(1) |
| K3-O7 | 2.904(3) | Os1-K3 | 3.912(2) |

Table 6. Bader's analysis of charges of atoms from cubic and triclinic $KOs_2O_6$ crystals.

| Cubic $KOs_2O_6$ | | Triclinic $KOs_2O_6$ | |
|---|---|---|---|
| atom | charge | atom | charge |
|  |  | K | 0.89+ |
|  |  | Os1 | 2.22+ |
| K | 0.92+ | Os2 | 2.38+ |
| Os | 2.43+ | O1 | 0.92- |
| O | 0.96- | O2 | 0.89- |
|  |  | O3 | 0.91- |
|  |  | O4 | 0.91- |
|  |  | O5 | 0.97- |
|  |  | O6 | 0.89- |